\documentclass{JHEP3}
\usepackage{amssymb}
\usepackage{epsfig}
\usepackage{graphicx}

\newcommand{\be}{\begin{equation}}
\newcommand{\ee}{\end{equation}}
\newcommand{\bea}{\begin{eqnarray}}
\newcommand{\eea}{\end{eqnarray}}

\def\Schw{Schwarzschild }
\def\({\left(} \def\){\right)}

\preprint{{\tt hep-th/0606156}}
\title{\center{On Black Fundamental Strings}}

\author{Amit Giveon and Dan Gorbonos\\
Racah Institute of Physics, The Hebrew University\\
Jerusalem 91904, Israel\\
E-mail: \email{giveon@phys.huji.ac.il}, \email{gdan@phys.huji.ac.il}}

\abstract{We study aspects of four dimensional black holes with
two electric charges, corresponding to fundamental strings with
generic momentum and winding on an internal circle.
The perturbative $\alpha'$ correction to such black holes and their
gravitational thermodynamics is obtained.}

\begin{document}

\section{Introduction}

Consider an excited fundamental string with mass $M$
in four dimensional flat space-time, and let the string
have momentum number $n$ and winding $w$ on an internal
circle of radius $R$ (for a review, see e.g.~\cite{Polchinski}).
The left- and right-handed momenta of such a string
are
\bea p_{L}=\frac{n}{R}-\frac{w\,R}{\alpha'}~,\nonumber\\
p_{R}=\frac{n}{R}+\frac{w\,R}{\alpha'}~,\label{plpr}\eea
where $\alpha'$ is the inverse string tension.
When the string interaction is strong enough
such strings form a four dimensional black hole
with a mass $M$ and two electric Neveu-Schwarz
charges $(p_L,p_R)$.

The leading order solutions of such black holes are well
known~\cite{HorPol}; it is reviewed in section 2.
The charged solutions can be constructed
from the Schwarzschild black hole by the following
procedure. First, one adds a fifth compactified
direction $x$ thus producing a uniform black string.
Then a boost is applied in this $x$ direction, which
after Kaluza-Klein reduction gives rise to a four dimensional
black hole with momentum charge $p_L=p_R=\frac{n}{R}$.
Performing a T-duality in the $x$ direction
gives instead a five dimensional black string winding $w$
times around the $x$ circle, and reducing to four dimensions
one obtains the black hole with winding charge.
Finally, a second boost of the black string in
the $x$ direction and reduction to four dimensions
generates the black hole with generic momentum
and winding charges $(p_L,p_R)$.
The gravitational thermodynamics of these solutions
was also inspected in~\cite{HorPol}, and we present
it in section 2.

In string theory, these solutions receive corrections
both in the string coupling and in $\alpha'$.
In this note we obtain the leading perturbative
$\alpha'$ correction
to four dimensional black holes which carry these generic
fundamental string charges $(p_L,p_R)$.

The leading order correction to the Schwarzschild solution
was derived by Callan-Myers-Perry (CMP) in~\cite{CMP},
and is reviewed in section 3.
We then apply the above procedure to the CMP solution.
In section 4, we perform the first boost thus generating
the $\alpha'$ corrections to the black hole with a momentum
charge. This was also discussed in~\cite{Itzhaki}.
We then investigate the effect of the correction on
some physical quantities such as the ADM mass and charges
as well as other thermodynamical quantities.

In section 5, we apply the T-duality and the second boost
to generate the $\alpha'$ corrections for general $(p_L,p_R)$.
Some simplification occurs in the ``heterotic'' case $p_L=0$
(where $x$ can be taken to be a chiral boson and embedded in
one of the right-handed extra bosonic directions)
which is the subject of section 6.

One of the main reasons to investigate $\alpha'$ corrections
to black fundamental strings is in order to learn
the way that string corrections affect the physics of black holes.
The results presented in this work have some consequences
which are discussed in section 7. In particular, they indicate
that $\alpha'$ corrections smear the horizon
in such a way that the entropy of black fundamental
strings is increased, as expected.

\section{4d Black holes with $(p_L,p_R)$ charges --
leading order in $\alpha'$}

The leading order solution describing a four dimensional black hole
with fundamental string charges $(p_L,p_R)$ (\ref{plpr})
in the string frame is~\cite{HorPol}
\be
ds^{2}=-\frac{\(1-\frac{2\,m}{r}\)}{\(1+\frac{2\,m\,\sinh^{2}(\alpha_{w})}{r}\)\,\(1+\frac{2\,m\,\sinh^{2}(\alpha_{n})}{r}\)}\,dt^{2}+\(1-\frac{2\,m}{r}\)^{-1}\,dr^{2}+r^{2}\,d\Omega_{2},
\label{hor-pol_metric} \ee
with a dilaton
\be
\phi(r)=\phi_0-\frac{1}{4}\ln\(1+\frac{2\,m\,
\sinh^{2}(\alpha_{w})}{r}\)
-\frac{1}{4}\ln\(1+\frac{2\,m\,\sinh^{2}(\alpha_{n})}{r}\),
\ee
where $\phi_0$ is a constant,
and two Abelian gauge fields
\bea
\emph{A}^{n}_{t}=\frac{m\,\sinh(2\,\alpha_{n})}
{r+2\,m\,\sinh^{2}(\alpha_{n})}~, \nonumber\\
\emph{A}^{w}_{t}=\frac{m\,\sinh(2\,\alpha_{w})}
{r+2\,m\,\sinh^{2}(\alpha_{w})}~.
\label{vec_potent}\eea
The boost parameters $\alpha_{w}$ and $\alpha_{n}$
(with $v=\tanh\alpha$ being a boost velocity)
correspond to the two boosts described in the introduction,
generating the winding and momentum charges, respectively.
$\emph{A}^{n}$ is the vector potential which is coupled to the
momentum of the fundamental string and $\emph{A}^{w}$
is coupled to its winding charge.
The horizon of the black hole is located at $r_h=2m$ for any
value of $\alpha_{n,w}$.
Here and below we set the four dimensional Newton constant to $G=1$.

The left and right moving momenta (\ref{plpr})
can be read from the asymptotic value of the
vector potentials:
\bea
p_{L}=\frac{m}{4}\(\sinh(2\,\alpha_{n})-\sinh(2\,\alpha_{w})\),\label{pppl}\\
p_{R}=\frac{m}{4}\(\sinh(2\,\alpha_{n})+\sinh(2\,\alpha_{w})\). \label{pppr}
\eea
The corresponding chemical potentials are\bea
\Phi_{L}=\frac{1}{2}\(\tanh(\alpha_{n})-\tanh(\alpha_{w})\), \label{phil}\\
\Phi_{R}=\frac{1}{2}\(\tanh(\alpha_{n})+\tanh(\alpha_{w})\); \label{phir}
 \eea
these are the electromagnetic potentials $A_t^{L,R}$ at the horizon,
where $(A_L,A_R)\equiv\frac{1}{2}(A^n-A^w,A^n+A^w)$.

The ADM mass is \be
M_{ADM}=\frac{m}{2}\(\cosh^{2}(\alpha_{w})+\cosh^{2}(\alpha_{n})\),
\label{admmass} \ee and the entropy is
 \be S=4\,\pi\,m^{2}\cosh(\alpha_{w})\cosh(\alpha_{n}). \label{smplpr} \ee
This expression can be interpreted as the \Schw entropy,
which is proportional to the area of the black string,
boosted twice along the additional fifth direction.
The inverse temperature $\beta=1/T$ is
 \be \beta=\(\frac{\partial\,S}
 {\partial\,M_{ADM}}\)_{p_{L},p_{R}}=8\,
 \pi\,m\cosh(\alpha_{w})\cosh(\alpha_{n}).
 \ee

The extremal limit is obtained by taking, say, $p_{R} \rightarrow M$.
This amounts to taking either $\alpha_n\to\infty$ and/or
$\alpha_w\to\infty$, as well as $m\to 0$,
such that $m(\exp(2\alpha_n)+\exp(2\alpha_w))$
is held fixed. In this limit the horizon is singular and,
in particular, $S \rightarrow 0$.

There are two special cases in which it is easy to write the
entropy as a function of the mass and the charges explicitly. When
$p \equiv p_{L}=p_{R}$, namely, no
winding charges: $\alpha_{w}=0$, using
(\ref{admmass}),(\ref{smplpr}),(\ref{pppl}) and (\ref{pppr}) we
get
\be
S=\frac{64\,\pi\,M_{ADM}^{2}\,(1-q^{2})^{\frac{3}{2}}}
{\(3+\sqrt{8\,q^{2}+1}\)^{2}}
\cdot\frac{\sqrt{1+2\,q^{2}+\sqrt{8\,q^{2}+1}}}{\sqrt{2}}
\label{explicit1},
\ee
where
\[q \equiv \frac{p}{M_{ADM}}.\]
The $\alpha'$ corrections to this case are
calculated in section 4.

When $p_{L}=0$, then $\alpha \equiv \alpha_{n}=\alpha_{w}$. Using
(\ref{admmass}),(\ref{smplpr}) and (\ref{pppr})
we obtain
\be
S
=4\,\pi\,\(M_{ADM}^{2}-p_{R}^{2}\).
\label{explicit2}
\ee
The $\alpha'$ corrections to this case are
calculated in section 6.

\section{The Callan-Myers-Perry solution}

We now review the higher derivative gravity corrections
to the Schwarzschild black hole.
The leading order correction to the low energy string effective action
$S_{eff}^0$ in $D$ dimensions gives rise to
\be S_{eff}^{\lambda}=
\frac{1}{16\,\pi\,G_{D}} \int d^{D}x
\,\sqrt{-g}\,e^{-2\,\phi}\,\left(R+4\,(\nabla\phi)^{2}
+\frac{\lambda}{2}\,R_{\mu\nu\rho\sigma}\,R^{\mu\nu\rho\sigma}\right),
\label{Seff}\ee
where $\lambda=\frac{\alpha'}{2},\,\frac{\alpha'}{4},\,0$ for
bosonic, heterotic and type II strings, respectively, and $G_{D}$
is the Newton constant in $D$ dimensions.

We shall concentrate on spherical symmetric solutions in
$D=4$. Hence, we take the following ansatz for a four
dimensional solution:
\be
ds^{2}=g_{tt}\, dt^{2}+g_{rr}\,dr^{2}+r^{2}\,d\Omega_{2}, \label{ansatz}
\ee
where $d\Omega_{2}$ is the metric on a unit $S^{2}$. The solution to
the equations of motion with the appropriate boundary conditions
(regularity at the horizon and asymptotic flatness as $r \rightarrow
\infty$) to first order in $\lambda$ (for $D=4$) ~\cite{CMP} is \bea
g_{tt}&=&-f\,\left(1+2\,\lambda
\,\mu(r)\right), \nonumber          \\
g_{rr}&=&f^{-1}\,\left( 1+2\,\lambda\,\epsilon(r)\right),
\label{Scwa corrected}\\
\phi&=&\phi_{0}+\lambda \,\varphi(r),\nonumber  \eea
where $\phi_{0}$ is the constant value of the dilaton in the zeroth
order solution in $\lambda$ and \bea
f&\equiv& 1-\frac{2\,m_{S}}{r}, \label{fff} \\
\mu(r)&=&-\left(\frac{23}{24\,m_{S}\,r}+\frac{11}{12\,r^{2}}
+\frac{m_{S}}{r^{3}}\right),\label{mmm}\\
\epsilon(r)&=&-\left(\frac{5\,m_{S}}{3\,r^{3}}
+\frac{7}{12\,r^{2}}+\frac{1}{24\,m_{S}\,r}\right)\label{eee},\\
\varphi(r)&=&-\left(\frac{2\,m_{S}}{3\,r^{3}}
+\frac{1}{2\,r^{2}}+\frac{1}{2\,m_{S}\,r}\right).
\eea
Note that this Schwarzschild-type metric depends on a single
mass parameter $m_{S}$. The horizon is located at $r_{h}=2\,m_{S}$.

We can write the
periodicity $\beta$ of the Euclidean time
using the surface gravity at the horizon
\be
\kappa=\left.\frac{1}{2}\,\frac{\left|\partial_{r}g_{tt}\right|}{\sqrt{-g_{tt}\,g_{rr}}}\right|_
{r=2\,m_{S}}=\frac{1}{4\,m_{S}}\left(1-\frac{11\,\lambda}{24\,m_{S}^{2}}\right),
\ee and then \be
\beta=\frac{2\,\pi}{\kappa}=8\,\pi\,m_{S}\,\left(1+\frac{11\,\lambda}{24\,m_{S}^{2}}\right).
\label{periodS} \ee

The action is apparently ``scheme'' dependent.
Different actions can be obtained by
redefinitions of the fields, which are in this case the
dilaton and the metric. Callan, Myers and Perry~\cite{CMP}
introduced a different scheme by using field redefinitions of order
$\lambda$ to eliminate the higher derivative dilaton terms that
appear after a conformal transformation to the Einstein frame,
\[g_{\alpha\beta}\rightarrow e^{2\,\phi}g_{\alpha\beta}.\]
 We will refer to the scheme adopted in the Einstein frame as ``the Einstein
scheme" and to the previous one as ``the String scheme.''

In the Einstein scheme the action (\ref{Seff}) becomes (for $D=4$)
\be S_{eff}=
\frac{1}{16\,\pi} \int d^{4}x
\,\sqrt{-g}\,\left(R-2\,(\nabla\phi)^{2}
+\frac{\lambda}{2}\,e^{-2\,\phi}\,R_{\mu\nu\rho\sigma}\,R^{\mu\nu\rho\sigma}
\right).
\ee
The corrections to the \Schw metric vanish in this scheme although
the correction to the dilaton remains unchanged. We will
denote the mass parameter here by $m_{E}$.\footnote{The mass parameter
$m$ in each scheme is defined such that the location of the horizon
is at $r_h=2m$, where the radial coordinate $r$ is normalized as imposed
by the ansatz (\ref{ansatz}).}  Since the solution
depends only on one parameter we can find the relation between
$m_{E}$ and $m_{S}$ to first order in $\lambda$ by comparing
scheme independent quantities like the periodicity $\beta$ or any
other thermodynamic quantity. Thermodynamic quantities and
identities are scheme independent. The periodicity $\beta$ in the
Einstein scheme is simply the periodicity of the \Schw solution
\be \beta=8\,\pi\,m_{E}. \ee Comparing to (\ref{periodS}), we
obtain the relation \be
m_{E}=m_{S}\,\left(1+\frac{11\,\lambda'}{6}\right),
\label{translation} \ee where we define \be \lambda' \equiv
\frac{\lambda}{r_{h}^{2}}=\frac{\lambda}{4\,m_{S}^{2}}. \ee
$\lambda'$ is the small dimensionless expansion parameter since
the $\alpha'$ corrections can be trusted only when the string
scale is small compared to the black hole size. The ADM mass in
the Einstein scheme \cite{CMP} is
\be M_{ADM}=m_{E}~,\ee
and can be translated to the string scheme using (\ref{translation}).

As a second example, we can apply this transformation to the free
energy that was calculated by Callan, Myers and Perry from the
Euclidean action in the Einstein scheme~\cite{CMP}, \be
\emph{F}_{E}=\frac{m_{E}}{2} \, \(1-2\,\lambda' \), \ee obtain
it in the string scheme as \be
\emph{F}_{S}=\frac{m_{S}}{2}\,\(1-\frac{\lambda'}{6}\), \label{free
energy}\ee and then derive the ADM mass in the string scheme using
the thermodynamic identity
\be M \equiv M_{ADM}=\frac{\partial \(\beta \, F_{S}\)}{\partial \beta}~.\ee
{}From now on we will denote the thermodynamical mass by $M$,
keeping in mind that it is the ADM mass.

The entropy in the string scheme is
 \be
 S=\beta\,\(M-F_{S}\)=
 4\,\pi\,m_{S}^{2}\(1+\frac{17}{3}\,\lambda'\), \label{CMP entropy}
\ee and it can be written in terms of the ADM mass, which is a scheme
invariant quantity, and $\lambda'$ as
 \be S=4\,\pi\,M^{2}\(1+2\,\lambda'\). \label{CMP
entropy inv}\ee

\section{The $\alpha'$ corrections to the $p_{L}=p_{R}$ solution}

In this section we calculate the $\alpha'$ corrections to the case
when $p \equiv p_{L}=p_{R}$ in (\ref{hor-pol_metric}), namely, no
winding charges: $w=\alpha_{w}=0$.
{}From now on we shall work in the string scheme (unless
otherwise specified).
As discussed in the introduction,
this solution is obtained from the
\Schw solution by adding a fifth spectator coordinate $x$ (to
create a black string), boosting along this direction and then
reducing to four dimensions. Thus we lift the four dimensional
solution (\ref{Scwa corrected}) to five dimensions and write
$g_{xx}$ to order $\lambda$ in the form
\[g_{xx}=1+\lambda\,\xi(r).\]
Then the equations of motion with the same appropriate boundary
conditions~\cite{CMP} as in the previous section
give $\xi = 0$ as the only solution.
Thus the boosted solution is \bea
\bar{g}_{tt}&=&\cosh^{2}(\alpha_{n}) \,g_{tt}+\sinh^{2}(\alpha_{n}),
\nonumber\\
\bar{g}_{xt}&=&\sinh(\alpha_{n})\,\cosh(\alpha_{n})\,(g_{tt}+1),
\label{boosted} \\
\bar{g}_{xx}&=&\sinh^{2}(\alpha_{n})\,g_{tt}+\cosh^{2}(\alpha_{n}),
 \nonumber\eea
 and after reduction to four dimensions we get to first order in
 $\lambda$:
\bea
\hat{g}_{tt}&=&\frac{g_{tt}}{\cosh^{2}(\alpha_{n})
+\sinh^{2}(\alpha_{n})\,g_{tt}}
=-\frac{f}{A}\(1+\frac{8\,m_{S}^{2}\,\lambda'\,\mu(r)\,
\cosh^{2}(\alpha_{n})}{A}\), \nonumber\\
\hat{\emph{A}_{t}^{n}}&=&\frac{\sinh(\alpha_{n})\,
\cosh(\alpha_{n})\,(g_{tt}+1)}{\cosh^{2}(\alpha_{n})
+\sinh^{2}(\alpha_{n})\,g_{tt}}=\frac{m_{S}\,
\sinh(2\,\alpha_{n})}{r\,A}\,\(1-\frac{4\,\lambda'\,\mu(r)\,m_{S}\,r\,f}{A}\),
\label{gtt in string}\\
e^{-2\,\hat{\phi}}&=&e^{-2\,\phi}\(\cosh^{2}(\alpha_{n})
+\sinh^{2}(\alpha_{n})\,g_{tt}\)^{\frac{1}{2}}
=e^{-2\,\phi}\,\sqrt{A}\,\(1-\frac{4\,\lambda'\,\mu(r)\,m_{S}^{2}\,
\sinh^{2}(\alpha_{n})\,f}{A}\)
\nonumber , \eea
 where
\be
 A \equiv 1+\frac{2\,m_{S}}{r}\,\sinh^{2}(\alpha_{n}).\label{aaa} \ee
The rest of the components of the four dimensional metric remain
unchanged.

 The charge $p$, which
 can be read from the coefficients of
 $r^{-1}$ at the asymptotic infinity, receives $\alpha'$ corrections:
 \be
 p=\frac{m_{S}}{4}\,\sinh(2\,\alpha_{n})\(1+\frac{23}{6}\lambda'\)
 \label{one p in s}.\ee
 On the other hand, the chemical potentials (\ref{phil},\ref{phir})
 are not affected by $\alpha'$
 corrections since the corrections to the vector potential
 vanish at the horizon. We will denote them by $\Phi \equiv \Phi_{L}=\Phi_{R}$.

 The periodicity of the Euclidean time is
 \be
\beta=8\,\pi\,m_{S}\,\cosh(\alpha_{n})\,\(1+\frac{11\,\lambda'}{6}\).
\label{period boosted once}
 \ee
 This result can be interpreted as the boosted inverse temperature
 of the \Schw solution with the $\alpha'$ correction
 (\ref{periodS}). Note that the boost parameter does not get
 $\alpha'$ corrections and hence it is kept in the same form in any scheme.

The ADM mass can be derived from the thermodynamic identity
\be M=\frac{\partial\( \beta\,F_S\)}{\partial \, \beta}+2\,p\,\Phi.
\label{follow}\ee
$\beta\,F$ is the value of the Euclidean action.
The integrand of the Euclidean action is invariant under the boost.
The only change is in the limits of integration in the $x$-$t$ plane.
Since the metric is static, the integration over the Euclidean time
gives only a multiplicative factor by its period $\beta$,
and the change in the size of the compactified direction $x$
is absorbed after the KK reduction by the four
dimensional Newton's constant.
Hence, the boost does not change the value of the free energy
and the result for the \Schw
case~(\ref{free energy}) is valid for the case considered here.
The ADM mass is~\footnote{It will also be derived from
the Einstein metric in subsection 4.1.}  \be
M=m_{S}\,\(1+\frac{11}{6}\lambda'\)+\frac{m_{S}\,\sinh^{2}(\alpha_{n})}{2}\(1+\frac{23}{6}\lambda'\).
\label{Mass in s}\ee
The charge $p$ and the mass $M$ are changed as a result of the
boost when $m_{S}$ and $\lambda'$ are held fixed. This way the
uncharged \Schw solution is transformed to a charged solution.
Note that the ratio
\[ \frac{\Delta \, M}{\Delta \, p}=\tanh(\alpha_{n}),\]
where $\Delta \,M$ and $\Delta \, p$ are the change in the mass and the charge,
is free of $\alpha'$ corrections.

One can obtain the corrected entropy from the thermodynamics as
well: \be
S=\beta\(M-F_S-2\,p\,\Phi\)=4\,\pi\,m_{S}^{2}\,\cosh(\alpha_{n})\,\(1+\frac{17}{3}\,\lambda'\).
\label{boosted entropy} \ee This result is the Callan-Myers-Perry
entropy (\ref{CMP entropy}) boosted with the parameter
$\alpha_{n}$. Before the reduction
the entropy is the product of the area of the black string
times a function of $\lambda'$, and when
we boost along the fifth dimension we expect to
get a boost factor of $\cosh(\alpha_{n})$ which reflects the
change in its size.

It is useful to write the parameters $m_{S}$ and $\alpha$ in terms
of $M$ and $p$, i.e. to invert the relations (\ref{one p in s})
and (\ref{Mass in s}):
\bea
\sinh^{2}(\alpha_{n})&=&\sinh^{2}(\alpha^{0}_{n})\,\(1-\frac{8\,\lambda'\,\cosh^{2}(\alpha^{0}_{n})}{2+3\,\sinh^{2}(\alpha^{0}_{n})}\)
,\label{conv1}
\\
m_{S}&=&m_{0}\,\(1-\lambda'\,\frac{22+21\,\sinh^{2}(\alpha^{0}_{n})}{6\,(2+3\,\sinh^{2}(\alpha^{0}_{n}))}\),
\label{conv2}
\eea where \bea \sinh^{2}(\alpha^{0}_{n})&=&\frac{4\,q^{2}-1+\sqrt{8\,q^{2}+1}}{2\,(1-q^{2})},\\
m_{0}&=&\frac{4\,M\,\(1-q^{2}\)}{3+\sqrt{8\,q^{2}+1}},\\
q &\equiv& \frac{p}{M}. \eea Using (\ref{conv1},\ref{conv2}) we can
express the entropy (\ref{boosted entropy}) in terms of scheme
invariant quantities -- the mass $M$ and the charge $p$: \be
S=S_{\lambda'=0}\(M,p\)\(1+2\,\lambda'\),\label{boosted entropy
inv}\ee where
\[S_{\lambda'=0}\(M,p\)=4\,\pi\,m_{0}^{2}\,\cosh(\alpha_{n}^{0})\]
as it is expressed explicitly in terms of $p$ and $M$ in
(\ref{explicit1}).
We see that the form of the perturbative $\lambda'$
correction to the entropy is a multiplicative factor which we
encountered in the Callan-Myers-Perry solution (\ref{CMP entropy
inv}).~\footnote{This result is general, as we shall see later.}

Now we would like to inspect at the {\it extremal} limit $q^2
\rightarrow 1$. In the type II superstring it corresponds to
$\frac{1}{2}$--BPS fundamental string states. Although in the
extremal limit $\sinh^{2}(\alpha_{n}) \rightarrow \infty$ and
$m_{S} \rightarrow 0$, we have \be m_{S}\,\sinh^{2}(\alpha_{n})
\rightarrow 2\,M\,\(1-\frac{23}{6}\,\lambda'\).\ee Note also that
$m_{S}\,\mu(r) \rightarrow -\frac{23}{24\,r}$. Substituting these
into the metric components (\ref{gtt in string}) we find that all
the $\lambda'$ corrections vanish, in agreement
with~\cite{HorTseyt} and ~\cite{Itzhaki}: \bea &&\hat{g}_{tt}
\rightarrow  -F(r),\qquad \hat{g}_{rr}  \rightarrow 1, \qquad
\hat{\emph{A}}_{t}^{n} \rightarrow 1-F(r),\qquad
\hat{\phi}\rightarrow\phi_0 + \frac{1}{4}\ln F(r),\nonumber\\
&&F^{-1}(r)\equiv1+{4M\over r}~.
 \eea
Note that even with the perturbative $\lambda'$ correction
the entropy $S \rightarrow 0$ and $\beta \rightarrow 0$
in the extremal limit.

\subsection{The Einstein scheme}

In this note, unless otherwise specified, we work in the string scheme.
Alternatively, in the Einstein scheme,
when we lift the \Schw solution up to five
dimensions there is a single function \cite{CMP,Itzhaki}
\be
g(r)=\frac{4\,m_{E}}{9\,r^{3}}+\frac{1}{3\,r^{2}}+\frac{1}{3\,m_{E}\,r}~,
\ee
which determines its $\alpha'$ corrections
in the form \bea g_{tt}^{E}&=&-f_E\,\(1-\lambda \, g(r)\),\\
g_{rr}^{E}&=&\frac{1}{f_E}\(1-\lambda \, g(r)\),\\
g_{xx}^{E}&=&1+2\,\lambda\,g(r), \\
\phi&=&\phi_{0}-\frac{3}{2} \lambda g(r), \eea where
\be f_E=1-\frac{2\,m_{E}}{r}~.\ee
 After a boost with the
parameter $\alpha_{n}$, reduction to four dimensions and
transformation to the Einstein frame, we get \bea
\hat{g}_{tt}^{E}&=&-\frac{f_E}{\sqrt{E}}\(1-\frac{3\,\lambda
\,g(r)\,f_E\,\sinh^{2}(\alpha_{n})}{2\,E}\),\label{gtte}\\
\hat{g}_{rr}^{E}&=&\frac{\sqrt{E}}{f_E}\(1+\frac{3\,\lambda
\,g(r)\,f_E\,\sinh^{2}(\alpha_{n})}{2\,E}\),\label{grre}\\
\hat{g}_{\theta\theta}^{E}&=&r^{2}\,\sqrt{E}\, \(1+\frac{3\,\lambda
\,g(r)\,f_E\,\sinh^{2}(\alpha_{n})}{2\,E}\),\\
\hat{\emph{A}}^{n\,E}_{t}&=&\frac{m_{E}\,
\sinh(2\,\alpha_{n})}{r+2\,m_{E}\,\sinh^{2}(\alpha_{n})}
\(1+\frac{3\,\lambda\,g(r)\,r^{2}\,f_E}{2\,m_{E}\,
(r+2\,m_{E}\,\sinh^{2}(\alpha_{n}))}\),\\
e^{-2\,\hat{\phi}}&=&e^{-2\,\phi}\,\sqrt{E}\,
\(1+\frac{\lambda\,g(r)\,\(2\,\cosh^{2}(\alpha_{n})
+f_E\,\sinh^{2}(\alpha_{n})\)}{2\,E}\),
\eea where
\be E \equiv 1+\frac{2\,m_{E}\,\sinh^{2}(\alpha_{n})}{r}~.
\ee
{}From the metric and the vector potential one can read the
corrections to the ADM mass, the period of the Euclidean time and
the charge, and get the same results that we obtained in the string
scheme (\ref{Mass in s},\ref{period boosted once},\ref{one p in
s}) written in terms of $m_{E}$.\footnote{For instance, by inspecting
the large $r$ behavior of the metric (\ref{gtte},\ref{grre}) one finds that
$-\hat{g}_{tt}^{E}=(\hat{g}_{rr}^{E})^{-1}=1-{2M\over r}+O({1\over r^2},
\lambda^2)$, with $M$ being the ADM mass derived previously in eq.
(\ref{Mass in s}) by using different methods.}

\section{The $\alpha'$ corrections for general $(p_{L},p_{R})$}

In this section we shall add winding charge by first T-dualizing
the solution of the previous section, in order
to change fundamental string's momentum into winding,
and then turning on another boost
to add momentum charge again.
So first, we apply T-duality in the $x$ direction
(for a review, see e.g.~\cite{GPR}) to the
solution in the previous section.
In the case without $\alpha'$ corrections T-duality reads:
\bea
\tilde{g}_{tt}&=&\bar{g}_{tt}-\frac{\bar{g}_{xt}^{2}}{\bar{g}_{xx}}
\label{abcd} \\
\tilde{g}_{xx}&=&\frac{1}{\bar{g}_{xx}}\\
\tilde{B}_{xt}&=&\frac{\bar{g}_{xt}}{\bar{g}_{xx}}\\
\tilde{\phi}&=&\phi-\frac{1}{2}\,\ln(\bar{g}_{xx})
\label{efgh}
 \eea
where $\bar{g}_{\mu\nu}$ is the boosted metric in five dimensions.
The rest of the fields and metric components do not change.

When one includes the $\alpha'$ corrections in the action the
T-duality rules are modified \cite{Meissner,Kaloper}.  In
\cite{Meissner} a different scheme is introduced in which the
$R_{\mu\nu\rho\sigma}\,R^{\mu\nu\rho\sigma}$ term in (\ref{Seff})
is transformed to a Gauss-Bonnet term \be R_{GB}^{2} \equiv
R_{\mu\nu\rho\sigma}\,
R^{\mu\nu\rho\sigma}-4\,R_{\mu\nu}\,R^{\mu\nu}+R^{2}~,\ee as well
as terms that involve derivatives of the scalar field. The full
action in this scheme is given in  eq. (4.5) of~\cite{Meissner}.
This ``Gauss-Bonnet scheme" is obtained from the string scheme
(\ref{Seff}) using the following field redefinitions~\footnote{The
field redefinitions are written for the case when the
antisymmetric tensor vanishes. Otherwise there are some additional
terms.} \bea g_{\mu\nu} &\rightarrow&
g_{\mu\nu}+2\,\lambda\,R_{\mu\nu} \\
\phi &\rightarrow& \phi +
\frac{\lambda}{4}\,R-\lambda\,\(\partial\,\phi\)^{2}.
 \eea
The Callan-Myers-Perry solution does not change under these field
redefinitions since all the terms of order $\alpha'$ are
identically zero.

In this scheme, Kaloper and Meissner \cite{Kaloper} obtained the
$\alpha'$ corrected rules written here for our particular case of a
diagonal metric which depends only on one coordinate $r$ and boosted
along the fifth additional direction $x$:
 \bea
\tilde{g}_{tt}&=&\bar{g}_{tt}-\frac{\bar{g}_{xt}^{2}}{\bar{g}_{xx}}
\label{gggtt}\\
\tilde{g}_{xx}&=&\frac{1}{\bar{g}_{xx}}\(1+\frac{\lambda\,\(\bar{g}_{xx,r}\)^{2}}{\bar{g}_{xx}^{2}\,g_{rr}}+\frac{\lambda\,\bar{g}_{xx}^{2}\,\(\partial_{r}\,V\)^{2}}{2\,g_{rr}\,g_{tt}}\)\\
\tilde{B}_{xt}&=&\frac{\bar{g}_{xt}}{\bar{g}_{xx}}-\frac{\lambda\,\partial_{r}\,V\,\bar{g}_{xx,r}}{g_{rr}\,\bar{g}_{xx}}\\
\tilde{\phi}&=&\phi+\frac{1}{4}\,\ln\(\frac{\tilde{g}_{xx}}{\bar{g}_{xx}}\),
\label{ppp}
\eea where
\[V \equiv \frac{\bar{g}_{xt}}{\bar{g}_{xx}}.\]
Note that in the particular case of the \Schw solution we have
$g_{rr}\,g_{tt}=-1$.

Substituting (\ref{Scwa corrected}) and (\ref{boosted})
with $\alpha_{w}$ instead of $\alpha_{n}$
in (\ref{gggtt}) -- (\ref{ppp}), we get \bea
\tilde{g}_{tt}&=&-\frac{f}{B}\(1+\frac{2\lambda\,\mu(r)\,\cosh^{2}(\alpha_{w})}{B}\)\\
\tilde{g}_{xx}&=&\frac{1}{B}\(1+\frac{2\,\lambda\,\sinh^{2}(\alpha_{w})}{B}\left[\mu(r)\,f+\zeta(r)\right]\)\\
\tilde{B}_{xt}&=&\frac{m_{S}\,\sinh(2\,\alpha_{w})}{r\,B}
\(1-\frac{\lambda\,\mu(r)\,f\,r}{m_{S}\,B}
-\frac{2\,\lambda\,m_{S}\,f\,\sinh^{2}(\alpha_{w})}{r^{3}\,B^{2}}\)
\label{last term}\\
\tilde{\phi}&=&\phi-\frac{1}{2}\,\ln(B)
+\frac{\lambda\,\sinh^{2}(\alpha_{w})}{2}\left[2\,\mu(r)\,f+\zeta(r)\right]
\eea
where $f,\mu(r)$ are given in (\ref{fff},\ref{mmm}), respectively, and
\be\zeta(r)=\frac{m_{S}^{2}}{r^{4}\,B}\(2\,f\,\sinh^{2}(\alpha_{w})
+\cosh^{2}(\alpha_{w})\)\label{zetar}~,\ee
\be B \equiv 1+\frac{2\,m_{S}\,\sinh^{2}(\alpha_{w})}{r}~.\ee
Now we perform the second boost in $x$ with a boost parameter
$\alpha_n$, and after reduction to four dimensions we find:
 \bea
\hat{g}_{tt}&=&\frac{\tilde{g}_{xx}\,\tilde{g}_{tt}}
{\cosh^{2}(\alpha_{n})\,\tilde{g}_{xx}+\sinh^{2}(\alpha_{n})\,\tilde{g}_{tt}}
\label{gtthat}\\
&=&-\frac{f}{A\,B}\left[1+\frac{2\,\lambda\,\mu(r)}{B}\,
\(\cosh^{2}(\alpha_{w})+\sinh^{2}(\alpha_{w})\,f\)
-\frac{2\,\lambda\,f\,\zeta(r)\,
\sinh^{2}(\alpha_{n})\,\sinh^{2}(\alpha_{w})}{A\,B}\right.
\nonumber\\
&-&\left. \frac{2\,\lambda\,\mu(r)\,f\(\sinh^{2}(\alpha_{w})
-\sinh^{2}(\alpha_{n})\)}{A\,B}\right]
\nonumber \\
\hat{g}_{rr}&=&f^{-1}\,\left( 1+2\,\lambda\,\epsilon(r)\right)\\
\hat{\emph{A}_{t}^{n}}&=&\frac{\sinh(\alpha_{n})\,\cosh(\alpha_{n})
\(\tilde{g}_{xx}+\tilde{g}_{tt}\)}{\cosh^{2}(\alpha_{n})\,
\tilde{g}_{xx}+\sinh^{2}(\alpha_{n})\,\tilde{g}_{tt}}\nonumber\\
&=&\frac{m_{S}\,\sinh(2\,\alpha_{n})}{r\,A}
\left[1-\frac{\lambda\,\mu(r)\,f\,r}{m_{S}\,B}
+\frac{\lambda\,r\,f\,\zeta(r)\,\sinh^{2}(\alpha_{w})}{m_{S}\,A\,B}
\right.\\
&-&\left.\frac{2\,\lambda\,\mu(r)\,f}{A\,B}\,\(\sinh^{2}(\alpha_{w})
-\sinh^{2}(\alpha_{n})\)\right] \nonumber\\
\hat{\emph{A}^{w}_{t}}&=&\tilde{B}_{xt} \label{awthat}\\
e^{-2\,\hat{\phi}}&=&e^{-2\,\phi}\,\sqrt{A\,B}\,
\left[1-\frac{2\,\lambda\,\mu(r)\,f\,\sinh^{2}(\alpha_{w})}{B}
+\frac{\lambda\,\mu(r)\,f}{A\,B}\(\sinh^{2}(\alpha_{w})
-\sinh^{2}(\alpha_{n})\)\right.\label{ephihat}\\
&+&\left.\frac{\lambda\,f\,\zeta(r)\,\sinh^{2}(\alpha_{w})\,
\sinh^{2}(\alpha_{n})}{A\,B}\right] \nonumber
\eea
where $\epsilon(r)$ and $A$ are given in (\ref{eee}) and (\ref{aaa}),
respectively.

A few comments are in order:

\begin{itemize}
\item
Note that the $\alpha'$ correction to T-duality in the metric and
dilaton manifests itself
in the terms that include the function  $\zeta(r)$ (\ref{zetar});
in the gauge fields, in addition to $\zeta$, the last term in
(\ref{last term}) is also due to the $\alpha'$ correction to T-duality.
\item
Those terms do not contribute to the asymptotic charges (since they behave
asymptotically as $r^{-4}$) nor to the extremal limit (discussed below).
\item
Excluding these terms in (\ref{gtthat}) -- (\ref{ephihat}),
the resulting background -- the one obtained by naive T-duality
(\ref{abcd})-(\ref{efgh}) -- is invariant under
$n\leftrightarrow w$.~\footnote{Even though it is not
manifest from the way we chose to present (\ref{gtthat}) -- (\ref{ephihat}).}
\end{itemize}

The horizon is located at $r_h=2m_S$, and
the Euclidean time has a period
\be
\beta=8\,\pi\,m_{S}\,\cosh(\alpha_{n})\,\cosh(\alpha_{w})\(1+\frac{11\,\lambda'}{6}\).
\ee
The charges can be read from the asymptotic behavior of the
vector potentials:
\bea
p_{L}=\frac{m_{S}}{4}\,\(\sinh(2\,\alpha_{n})
-\sinh(2\,\alpha_{w})\)\(1+\frac{23}{6}\lambda'\) \label{general pl}\\
p_{R}=\frac{m_{S}}{4}\,\(\sinh(2\,\alpha_{n})
+\sinh(2\,\alpha_{w})\)\(1+\frac{23}{6}\lambda'\)\label{general pr}
\eea
and the chemical potentials (\ref{phil},\ref{phir}) again do not
receive $\alpha'$ corrections.

Since the integrand of the Euclidean action is invariant under
T-duality, and following the discussion after eq. (\ref{follow}),
the free energy (\ref{free energy}) is valid in the present case as well.
The mass is thus
\bea M&=& \frac{\partial\(
\beta\,F_{S}\)}{\partial \,
\beta}+p_{L}\,\Phi_{L}+p_{R}\,\Phi_{R} \nonumber\\
&=&m_{S}\,\(1+\frac{11}{6}\lambda'\)+\frac{m_{S}\,
(\sinh^{2}(\alpha_{n})+\sinh^{2}(\alpha_{w}))}{2}\(1+\frac{23}{6}\lambda'\),
\label{general m}\eea
and the entropy is
 \be
 S=\beta\,\(M-F_{S}-p_{L}\,\Phi_{L}-p_{R}\,\Phi_{R}\)
 =4\,\pi\,m_{S}^{2}\,\cosh(\alpha_{n})\,\cosh(\alpha_{w})
 \(1+\frac{17}{3}\,\lambda'\). \label{pfe}
 \ee
One can show that
\be
S=S_{\lambda'=0}(M;p_L,p_R)(1+2\lambda')~, \label{full entropy}
\ee
where $S_{\lambda'=0}(M;p_L,p_R)$ is the value of the leading order
entropy -- the Bekenstein-Hawking one -- of a black fundamental string
with mass $M$ and charges $(p_L,p_R)$.
The derivation of (\ref{full entropy})
for the general case is sketched in the appendix,
while the particularly simple ``heterotic''
case $p_L=0$ is presented in the next section,
where we also discuss the extremal limit.

\section{The $\alpha'$ corrections to the $p_{L}=0$ solution}

In the case when $\alpha \equiv \alpha_{n}= \alpha_{w}$  the
expressions simplify significantly: \bea
\hat{g}_{tt}&=&-\frac{f}{A^{2}}\left[1+\frac{2\,\lambda\,\mu(r)}{A}\,\(\cosh^{2}(\alpha)+\sinh^{2}(\alpha)\,f\)-\frac{2\,\lambda\,f\,\zeta(r)\,
\sinh^{4}(\alpha)}{A^{2}}\right]
\nonumber\\
\hat{g}_{rr}&=&f^{-1}\,\left( 1+2\,\lambda\,\epsilon(r)\right) \nonumber\\
 \hat{\emph{A}}^{n}_{t}&=&\frac{m_{S}\,\sinh(2\,\alpha)}{r\,A}\left[1-\frac{\lambda\,\mu(r)\,f\,r}{m_{S}\,A}+\frac{\lambda\,r\,f\,\zeta(r)\,\sinh^{2}(\alpha)}{m_{S}\,A^{2}}\right]
 \\
\hat{\emph{A}}^{w}_{t}&=&\frac{m_{S}\,\sinh(2\,\alpha)}
{r\,A}\left[1-\frac{\lambda\,\mu(r)\,f\,r}{m_{S}\,A}
-\frac{2\,\lambda\,m_{S}\,f\,\sinh^{2}(\alpha)}{r^{3}\,A^{2}}\right]
\nonumber\\
 e^{-2\,\hat{\phi}}&=&e^{-2\,\phi}\,A\,
 \left[1-\frac{2\,\lambda\,\mu(r)\,f\,\sinh^{2}(\alpha)}{A}
 +\frac{\lambda\,f\,\zeta(r)\,\sinh^{4}(\alpha)}{A^{2}}\right].\nonumber
\eea
Recall that $\hat{\emph{A}}^{n}_{t} \neq
\hat{\emph{A}}^{w}_{t}$ due to the modified T-duality rules.

The charges are \bea
    p_{L}&=&0~,\\
    p_{R}&=&\frac{m_{S}}{2}\,\sinh(2\,\alpha)\(1+\frac{23}{6}\lambda'\),
    \label{pR in s=}
\eea and the mass is \be
M=m_{S}\,\(1+\frac{11}{6}\lambda'\)
+m_{S}\,\sinh^{2}(\alpha)\(1+\frac{23}{6}\lambda'\).
\label{Mass in s=} \ee
{}From (\ref{pR in s=}) and (\ref{Mass in s=}) we can write $\alpha$
and $m_{S}$ in terms of $p_{R}$ and $M$:
\bea
m_{S}&=&M\,(1-q^{2})\,\left[1+(2\,q^{2}-\frac{11}{6})\lambda'\right],
\\
\sinh^{2}(\alpha)&=&\frac{q^{2}}{1-q^{2}}\,\(1-4\,\lambda'\), \eea
where
\be q\equiv \frac{p_{R}}{M}~.\ee
The entropy (\ref{full entropy}) in terms of $p_{R}$ and $M$ is
\be S=S_{\lambda'=0}\(M,p_{R}\)\,\(1+2\,\lambda'\),
\label{sbha} \ee
where
\be
S_{\lambda'=0}\(M,p_{R}\) =4\,\pi\,\(M^{2}-p_{R}^{2}\).
\label{smpr}\ee
In the {\it extremal} limit $q^2 \rightarrow 1$ one obtains
\be m_{S}\,\sinh^{2}(\alpha) \rightarrow M\,\(1-\frac{23}{6}\lambda'\)~,\ee
and the background becomes:
\bea
&&\hat{g}_{tt} \rightarrow -G(r)^2,\qquad
\hat{g}_{rr} \rightarrow 1, \qquad
\hat{A}^{n}_{t}, \hat{A}^{w}_{t}\rightarrow 1-G(r),\qquad
\hat{\phi}\rightarrow \phi_0 +\frac{1}{2}\ln G(r),
\nonumber \\
&&G^{-1}(r) \equiv 1+{2M\over r}~.\label{phiexte} \eea In the type
II superstring this solution corresponds to $\frac{1}{4}$--BPS
fundamental string states.

Note that in the extremal limit there are no perturbative $\lambda'$
corrections, in agreement with \cite{HorTseyt}, thus in particular the
horizon becomes singular and $S \rightarrow 0$.
Finally, for {\it generic} left-handed momentum $p_L$,
in the extremal limit $p_R^2\to M^2$ we also find that there are no
perturbative $\alpha'$ corrections and, in particular, $S\to 0$;
this is sketched in the appendix.

\section{Discussion}

In this work we studied corrections to four dimensioanl
{\it non}-extremal black hole solutions of string theory
with fundamental string momentum and winding charges $n,w$,
in the presence of higher derivative corrections
of the form $\lambda R^2$, perturbatively in $\lambda$.
We found (\ref{full entropy}) that the Bekenstein-Hawking
entropy is increased by an overall factor $1+{2\lambda\over r_h^2}$,
where $r_h$ is the radius of the black hole.

The perturbative analysis in $\lambda$ is reliable
for large black holes.
The investigation of the physics of small black holes
requires the exact solution in $\alpha'$.
Nevertheless, it was recently shown that {\it exact}
solutions in $\lambda$ to certain classes of $\lambda R^2$
theories is sufficient to obtain exact properties of small
black holes. For instance, higher derivative corrections
of the form $\lambda R^2$ to supersymmetric $N=2$ black holes
in four dimensions, stretch the singular horizon discussed in
section 2 to an $AdS_2\times S^2$ whose {\it finite}
radii as well as the value of the dilaton are fixed in terms
of the charges at their attractor point (for a recent review,
see e.g. \cite{Mohaupt}).
The power of supersymmetry allows to prove that this near horizon
geometry is exact in $\alpha'$.
The Wald formula~\footnote{The ``Euclidean procedure'' for obtaining
the black hole entropy, which we have used in this work, must give
\cite{Wald} the same results as obtained by Wald's formula.}
in higher derivative gravity \cite{Wald,IW}
and the supersymmetric attractor equations allow to compute the
exact entropy. Remarkably, for a black heterotic string
with electric charges $(p_L,p_R=\pm M)$
the Wald formula produces the entropy of free fundamental strings
with the same charges~\cite{Dabholkar}.

Recently, Sen studied the effect of the Gauss-Bonnet term on the
entropy of fundamental heterotic strings, by inserting an $AdS_2\times S^2$
ansatz into the equations of motion and applying the Wald formula.
Surprisingly, he found \cite{Sen} the same background obtained via the
attractor mechanism and consequently, for electrically charged cases,
the entropy of free heterotic strings with charges $(p_L,p_R=\pm M)$.

It is not clear though how to ``glue'' the
near horizon $AdS_2\times S^2$ background to an asymptotically flat
space-time. At present, the exact solution to the $\lambda R^2$ theory
is not known explicitly even in the uncharged case.
Once such a solution is found, we can apply
to it the techniques used in this work in order to charge
the black hole and, in particular, take the extremal limit.
It would then may be possible to
examine how the singular horizon is stretched, and if and how
the near horizon background is connected to
an asymptotically flat space-time for small black holes.

Finally, let us mention that
recent studies of two dimensional
black holes \cite{GKRS,Kutasov,GK}
indicate that non-perturbative $\alpha'$ effects of the exact
Conformal Field Theory
are dominant near the horizon of small black holes.
In particular,
the two dimensional extremal black hole obtained as a solution
to the perturbative equations of motion is asymptotically flat,
but the {\it exact} CFT background is actually $AdS_2$ \cite{GS}.
Hence, similar phenomena may
occur in four dimensional black holes as well,
as proposed in \cite{Kutasov,GK}.
One way or another, the exact solution in $\lambda$
is required to shed light on these interesting issues.

\bigskip
\noindent{\bf Note Added:}
After this work was completed,
the paper \cite{Kats:2006xp} appeared,
where the hypothesis \cite{Arkani-Hamed:2006dz}
that higher derivative corrections always make
extremal non-supersymmetric black holes
lighter than the classical bound was investigated.
In our examples the horizon becomes singular in the extremal limit
and thus, as discussed above,
perturbative $\alpha'$ expansion is not reliable.
Yet, the results (\ref{general pl})--(\ref{general m})
imply that, say,
$${M\over p_R}=
{\cosh^2(\alpha_n)+\cosh^2(\alpha_w)-4\lambda'\over
\sinh(\alpha_n)\cosh(\alpha_n)+\sinh(\alpha_w)\cosh(\alpha_w)}~,$$
and thus $\alpha'$ corrections decrease the mass/charge ratio
for {\it any} electric charge $(p_L,p_R)$.

\bigskip
\noindent{\bf Acknowledgements:}
We thank Barak Kol and the referee for their
comments on the manuscript.
DG thanks Vadim Asnin, Henriette
Elvang, Gary Horowitz, Dan Israel, Barak Kol, Ya'akov
Mandelbaum, Rob Myres, Simon Ross and Misha Smolkin  for
discussions, and the KITP in Santa-Barbara for hospitality.
This research is supported in part by the BSF -- American-Israel
Bi-National Science Foundation, the
Israel Science Foundation (grant No.1398/04),
the EU grant MRTN-CT-2004-512194,
and by a grant of DIP (H.52).

\appendix
\section{The entropy and the extremal limit for general $(p_{L},p_{R})$}

First we show how to obtain eq. (\ref{full entropy}).
Like in the simple cases where $p_{L}=p_{R}$
or $p_{L}=0$ we would like to write the parameters $m_{S}$,
$\alpha_{n}$ and $\alpha_{w}$ in terms of $M,$ $p_{L}$ and $p_{R}$.
Actually, it is not necessary to obtain the inversion of
(\ref{general pl}), (\ref{general pr}) and (\ref{general m}) explicitly.
Instead, let us write:
\bea
\sinh(\alpha_{n})&=&\sinh(\alpha_{n}^{0})\(1+a_{n}\,\lambda'\),\\
\sinh(\alpha_{w})&=&\sinh(\alpha_{w}^{0})\(1+a_{w}\,\lambda'\),\\
m_S&=&m_{0}\(1+b\,\lambda'\), \eea
where we denote by
$\sinh(\alpha_{n}^{0})$, $\sinh(\alpha_{w}^{0})$ and $m_{0}$ the
inversion of (\ref{general pl}), (\ref{general pr}) and (\ref{general m})
in the zeroth order (when $\lambda'=0$). Substitution into
(\ref{general pl}), (\ref{general pr}) and (\ref{general m}) gives
\bea
a_{n}&=&-\frac{b+\frac{23}{6}}{1+\tanh^{2}(\alpha_{n}^{0})}~,\\
a_{w}&=&-\frac{b+\frac{23}{6}}{1+\tanh^{2}(\alpha_{w}^{0})}~,\\
b+\frac{23}{6}&=&\frac{4}{\frac{1}{1+\tanh^{2}(\alpha_{n}^{0})}
+\frac{1}{1+\tanh^{2}(\alpha_{w}^{0})}}~.
\eea
Thus, substituting
\bea
\cosh(\alpha_{n})=\cosh(\alpha^{0}_{n})\,
\(1+a_{n}\,\tanh^{2}(\alpha_{n}^{0})\,\lambda'\),\\
\cosh(\alpha_{w})=\cosh(\alpha^{0}_{w})\,
\(1+a_{w}\,\tanh^{2}(\alpha_{w}^{0})\,\lambda'\),
\eea
and $m_S$ in (\ref{pfe}), we obtain
\be
S=4\,\pi\,m_{0}^{2}\,\cosh(\alpha^{0}_{n})\,
\cosh(\alpha^{0}_{w})\,\(1+2\,\lambda'\),\label{the entropy}
\ee
which is eq. (\ref{full entropy}).

Finally,
the extremal limit $p_{R}^{2} \rightarrow M^{2}$ is obtained when
we take $\alpha_{n}^{0},\alpha_{w}^{0} \rightarrow \infty$,
$m_{0} \rightarrow 0$ and keep $m_{0}\sinh^{2}(\alpha_{n}^{0})$
and $m_{0}\sinh^{2}(\alpha_{w}^{0})$ fixed. In this limit
$S\to 0$, and
\bea
a_{n},a_{w} &\rightarrow& -2~,\\
b &\rightarrow& \frac{1}{6}~, \eea
which imply
\bea
m_{S}\,\sinh^{2}(\alpha_{n}) &\rightarrow&
m_{0}\,\sinh^{2}(\alpha^{0}_{n})\(1-\frac{23}{6}\,\lambda'\) ,\nonumber\\
m_{S}\,\sinh^{2}(\alpha_{w}) &\rightarrow&
m_{0}\,\sinh^{2}(\alpha^{0}_{w})\(1-\frac{23}{6}\,\lambda'\).
\nonumber \eea
Substituting these into (\ref{gtthat}) -- (\ref{ephihat}),
one finds that the $\lambda'$ corrections vanish.

\end{document}